\documentclass[runningheads]{llncs}
\usepackage{amsmath,amssymb,amsfonts}
\usepackage{algorithmic}
\usepackage{graphicx}
\usepackage{textcomp}
\usepackage{xcolor}
\usepackage{blindtext}
\usepackage{tabto,lipsum,soul} 
\usepackage{multirow}
\usepackage{float}
\usepackage{rotating}
\usepackage{pgf-pie}
\usepackage{wrapfig}
\usepackage{subfig}
\usepackage{tikz}
\usepackage{enumerate}
\usepackage[shortlabels]{enumitem}
\usepackage{comment}
\usepackage{adjustbox}
\DeclareUnicodeCharacter{0308}{\textasciicircum}

\usepackage[utf8]{inputenc} 

\usepackage[T1]{fontenc}
\usepackage{pgfplots}

\pgfplotsset{
  compat=newest,
  xlabel near ticks,
  ylabel near ticks
}

\usepackage[
]{biblatex}
\begin{document}
\title{A meta-analytical comparison of Naive Bayes and Random Forest for software defect prediction}
\titlerunning{Meta-analysis for software defect prediction}
\author{Ch Muhammad Awais\inst{1} \and
Wei Gu\inst{1} \and
Gcinizwe Dlamini\inst{1}\and
Zamira Kholmatova\inst{1}\and
Giancarlo Succi\inst{2}
}
\authorrunning{A. Ch Muhammad et al.}
\institute{Innopolis University, Russia\inst{1}, Universit\`{a} di Bologna, Italy\inst{2}
\email{\{c.awais,g.wei,g.dlamini,z.kholmatova\}@innopolis.university, g.succi@unibo.it}}
\maketitle              
\begin{abstract}
Is there a statistical difference between Naive Bayes and Random Forest in terms of recall, f-measure, and precision for predicting software defects? By utilizing systematic literature review and meta-analysis, we are answering this question. We conducted a systematic literature review by establishing criteria to search and choose papers, resulting in five studies. After that, using the meta-data and forest-plots of five chosen papers, we conducted a meta-analysis to compare the two models. The results have shown that there is no significant statistical evidence that Naive Bayes perform differently from Random Forest in terms of recall, f-measure, and precision.

\end{abstract}
\section{Introduction}
\label{sec:introduction}
A defect that causes software to behave unexpectedly in a way that does not meet the actual requirements is known as a software defect \cite{devnani1998modeling}. Software defects can be heavily influential, which gives rise to disasters.
Nowadays, software systems are increasing steadily in size and complexity \cite{zuse2019software}. Industry has developed sophisticated software testing methodologies \cite{lewis2017software}. To ensure quality, newly developed or modified engineering products are often subjected to rigorous testing. 

However, the nature of software testing is time-consuming and resource-hungry, typically the software testing process takes up approximately 40\%-50\% of the total development resources, 30\% of the total effort and 50\%-60\% of the total cost of software development \cite{kumar2016impacts}. Over the past two decades, a variety of machine learning approaches have been employed to detect software defects in an endeavor to reduce, to minimize cost and automate the task of software testing \cite{aleem2015benchmarking}. 

In a typical workflow for building an ML defect prediction model, the data usually comes from the version control system and contains the source code and commit messages. Accordingly, an instance can be labelled as defective or non-defective. The instances in the data set are then constructed based on the acquisition of tags and other messages in the version control system, and a defect prediction model can be built by using a set of training instances. Finally, the prediction model can classify whether a new test instance has a defect or not \cite{li2018progress}. Our focus is on comparing prediction methods based on the following two ML models: Naive Bayes (NB) Random forest (RF) \cite{jacob2015improved} 

In this paper, we present a systematic literature review and meta-analysis of two machine learning methods for software defect prediction. Our motivation is to help researchers and practitioners in the field of software defect prediction to build a better understanding of these two algorithms that have been applied from a very early stage. The goal is to find out if there is any significance performance difference between NB and RF on software defect prediction in terms of precision, recall and f-measure.

The remaining sections of this paper are organized as follows: In section \ref{sec:relatedWorks}, related works are outlined. Section \ref{sec:methodology} describes the methodology, systematic literature review and meta-analysis. Section \ref{sec:results} presents the result and the discussion. Section \ref{sec:conclusion} outlines the conclusion.

\section{Related Works}
\label{sec:relatedWorks}
Software defect prediction is executed using some software related data, that consists of information about the code for software development, i.e. lines of code, number of methods, inheritance etc. In the paper \cite{he2015simplmetricSet}, researchers have discussed the effects of the metric set on software defect prediction, which summarizes that, if we precisely simplify metrics/features of our dataset, then the simplest model will perform better for software defect prediction, meaning that the feature selection plays an important role in software defect prediction.

Machine learning models \cite{venteka2005emperical} are used for performing software defect prediction. The researchers \cite{son2019depMappings} discussed the techniques for building an ideal model capable of detecting unseen defects in software. 
Software defect prediction models only focus on training on a specific domain of data, i.e., open-source projects, 
whereas an ideal model is trained on commercial projects. Moreover, preprocessing (i.e feature selection, multi-co-linearity) of the dataset for training, can improve the performance of the defect prediction model significantly \cite{son2019depMappings}. 

Meta-analysis is the procedure of combining different studies for finding out whether the experiments performed in different studies \cite{leandro2008meta} follow the same distribution or not. It helps in validating the results in different studies. The limitations of meta-analysis for defect prediction \cite{miller2000applying}, resulted in a guide for future researchers for meta-analytical studies on software defect prediction.

Software defect prediction can vary based on the knowledge and skills of the researcher \cite{martin2016researcherBias}. In \cite{martin2016researcherBias}, a meta-analytical experiment helped in finding bias in the results of software defect prediction, comparing 4 different groups (Classifier's effect, Dataset effect, effect of metrics and effect of researchers). The results showed that the effect of the classifier on the result was smaller than the effect of the researcher on the study result.

Soe et al. \cite{soe2018NB} performed experiments on 12 different datasets, in nine datasets RF performed better than NB with a marginal difference, in one the accuracy was the same, and in one NB outperformed RF with a difference of 0.07\%. The accuracy of NB on the PC3 dataset was i.e 45.8\%, while RF achieved 89.76\%, which made a huge difference of 43.89\%, because of these differences the researchers stated that Random Forest is better. We are testing the claim of \cite{soe2018NB}, using meta-analysis and systematic literature review.

\section{Methodology}
\label{sec:methodology}

\begin{figure}[ht]
    \centering
        \includegraphics[width=11.0cm, height=2.0cm]{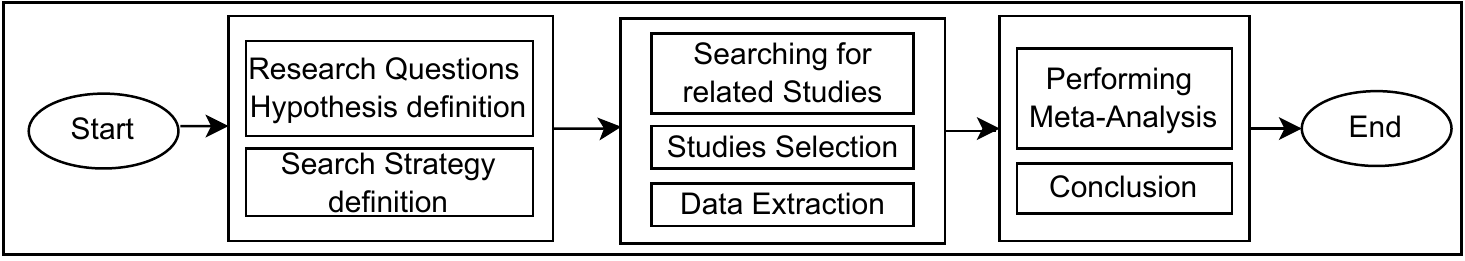}
        \caption{Methodology Overview}
        \label{fig:reviewMethodology}
\end{figure}

Our proposed methodology is presented in Fig. \ref{fig:reviewMethodology}. It contains two main stages: A) Literature Review and B) Meta-analysis. The following sections outline the details about each stage.

\subsection{Literature Review}
To test our aforementioned hypothesis using a meta-analytical approach and address the research questions, we conducted a systematic literature review. 

\subsubsection{Research Questions and hypothesis definition:} Research questions (RQ) helps in steering research to valuable and constructive conclusions \cite{j2009develop}. We apply PICOC(Population, Intervention, Comparison, Outcomes, and Context) \cite{petticrew2008picoc} approach to produce five main research questions, which drive the analysis application of machine learning techniques (i.e naive Bayes, Random forest) in software defect prediction.  
\begin{enumerate}[start=1, label={\bfseries RQ\arabic*.}, wide = 0pt, leftmargin = 3em]
    \item Which repositories are popular for software defect prediction?
    \item What kinds of datasets are the most commonly utilized in prediction of software defects?
    \item What are the programming languages which commonly in datasets for software defect prediction?
    \item What are the common methods that are used to predict software defects?
    \item What comparison metrics are commonly used for comparing the software defect prediction performance?
\end{enumerate}

\subsubsection{Search strategy:}
Before beginning the search, it is important to choose a suitable combination of databases to enhance the chances of discovering extremely relevant articles. Digital libraries are the databases where the papers are published, it is an important step to decide which libraries to choose. To have the broadest set of studies possible, it is important to search for literature databases that offer a broad perspective on the field. We used Google Scholar (GS), Research Gate (RG), Springer (SP) as search databases.

Based on our research questions we formulated the following query string: 
(software OR applicati* OR systems) AND (fault* OR defect* OR quality OR error-prone) AND (predict* OR prone* OR probability OR assess* OR detect* OR estimat* OR classificat*) AND (random* OR forest) AND (naive OR bayes)

The human behavior may differ based on their own selection criteria, so the search string was adjusted based on the trends being used for searches, also the original one was kept. After the search string was adjusted to suit the specific requirements of each database, the databases were searched by title, keyword, and abstract. Search results limited to publications published from 1999 to 2021 were retrieved. Only English journal papers and conference proceedings were included.

\subsubsection{Study selection:}
We will discuss the tools and techniques used for selecting the studies for performing the meta-analytical study. Also in this section we will explain the process of filtering retrieved studies requires outlining inclusion and exclusion criteria. To eliminate bias in the systematic literature review procedure, we defined the inclusion and exclusion criteria using PICOC approach in advance.
\begin{itemize}
    \item \textbf{Inclusion Criteria}:
    \item[] We included studies that
    \begin{itemize}
        \item are discussing NB and RF models in software defect prediction.
        \item are with publicly available datasets for software defect prediction.
        \item benchmarked performance with appropriate metrics for comparison.
    \end{itemize}
    \item \textbf{Exclusion criteria}
    \begin{itemize}
        \item Studies that do not include our research objects.
        \item Studies with unpublished datasets.
        \item Studies that are with no validation of experimental results or experimental process.
        \item Non-peer-reviewed studies.
        \item Systematic review studies.
    \end{itemize}
\end{itemize}

Initially, the literature reviews and non peer-reviewed papers were removed. We created groups in Mendeley\cite{zaugg2011mendeley} to collaborate on the screening of studies. Two reviewers independently reviewed all the works, decisions on whether to screen a study were made according to established exclusion and inclusion criteria. When reviewers disagree, a discussion will occur and an explanation for screening will be provided. If there is uncertainty for the final decision, then the literature will be included for analysis.

\subsubsection{Selection results:}
\label{sec:searchselection}
We have collected 62 research studies based on our search query from different databases. On this stage we applied screening based on title(Removal of papers with unrelated titles) and removal of duplicates.

\subsubsection{Assessment criteria:}
\label{sec:assessmentCriteria}
We utilized a brief checklist inspired by Ming's \cite{li2012sample} work to determine whether the study provides adequate contextual and methodological information, to interpret if the study is sufficient to answer the research questions. After a pilot test, we fine-tuned the criteria for our research questions using the following different stages.
\begin{itemize}
    \item \textbf{Stage 1:} Context Criteria
        \begin{itemize}
            \item The study's objective and domain should be clearly stated.
            \item The programming language for benchmarking must be specified.
            \item The source information of publicly available dataset should be provided.
        \end{itemize}
    \item \textbf{Stage 2:} Model building Criteria
        \begin{itemize}
            \item The features used for training(e.g, software metrics), must be explicit.
            \item The dependent variable predicted by the models should be stated clearly.
            \item The granularity of the dependent variable should be reported, i.e., whether the predicted outcome is a statistic on LOC, modules, files, or packages.
        \end{itemize}
    \item \textbf{Stage 3:} Data Criteria
        \begin{itemize}
            \item There should be a report or reference about data collection in order to have a confidence on the dataset.
            \item The data transformation or pre-processing techniques should be clearly stated in the study.
        \end{itemize}
    \item \textbf{Stage 4:} Predictive performance Criteria
        \begin{itemize}
            \item The prediction model must be tested on unseen data after training.
        \end{itemize}
\end{itemize}

\subsubsection{Assessment results:}
We performed full-text analysis on each study to keep paper, which discussing Naive Bayes and Random Forests. We excluded the studies which were not discussing the specific metrics(i.e. recall, precision, f-measure).

\subsubsection{Data Extraction:}
We carried out a pilot test on three randomized studies to observe if data extraction meets the needs of analysis. Two group members separately designed the table headers, after the pilot test, they discussed and combined the two tables to ensure the structure consistency. The data was extracted according to the established table structure, the consistency of data was compared to ensure that no mistakes were made in the extraction process.

\subsection{Meta-Analysis}
\label{sec:metaanalysis}
In meta-analysis, firstly, we find out the outcome of a study, and based on this outcome, statistics are calculated for each study. Secondly, we combine the statistics of all studies to estimate the summary by using the weighted average of each study. Based on the number of experiments, and the publishing year, the study is given weights or other factors, i.e. study with fewer experiments will get a lower weight etc. \cite{c2009systematic.ch1}. To understand meta-analysis, we need to define a few terms, which are as follows:
\begin{itemize}
    \item \textbf{Effect Size:} The measure of an outcome to validate/invalidate the hypothesis. We used mean-difference because our study is based on the recall score.
    \item \textbf{Forest Plot:} The Forest plot indicates the variability in the results, based on the forest plot we decide the presence of heterogeneity.
    \item \textbf{Heterogeneity:} The validation criterion to decide whether to employ meta-analysis, i.e. high heterogeneity means the studies are not linking up, and it leads to halt in meta-analysis. In our case, low heterogeneity can be later visualized in the forest plots.
    \item \textbf{Effect Model:} We utilized two effects models, i.e. fixed and random. We will discuss them in the next section.
\end{itemize}
\subsubsection{Fixed vs Random Effect Models}
Fixed effect model is used for best estimation of the effect size, it is presumed that all studies have a single effect. Whereas random effect model is used for average effect, in random effect it is assumed that each study has a different effect \cite{c2009systematic.ch1}. In order to estimate the individual effects of a assessment criterion and the difference between the experimental and control groups, we calculate the effect size.
The below formulas from \cite{intro_to_Meta_book_2009.ch16} are used for calculation of the Effect Size $g$ and its standard deviation $SE_g$:

\begin{equation}
g = J \times d
\end{equation}

\noindent $j$ is Hedge's correction factor: $J=1-\frac{3}{4 d f-1}$ and $df=NumberofStudies-1$
\noindent $d$ is Cohen's standardized difference of the sample means:
\begin{equation}
\label{eq_d}
d=\frac{\bar{X}_{1}-\bar{X}_{2}}{S_{\text {within }}}
\end{equation}

\noindent Here $X_1$ and $X_2$ are the means of the control group (Naive Bayes) and the experimental group (Random Forest) respectively, with $S_{\text {within}}$ being the pooled within-groups standard deviation (with an assumption that $\sigma_1 = \sigma_2$):
\begin{equation}
\label{eq_S_within}
S_{\text {within }}=\sqrt{\frac{\left(n_{1}-1\right) S_{1}^{2}+\left(n_{2}-1\right) S_{2}^{2}}{n_{1}+n_{2}-2}}
\end{equation}
where $S_1$,$S_2$ are the standard deviation of control group and experimental group and $n_1$,$n_2$ are the number of studies of control group and experimental group respectively.
\newline\noindent The variance and standard deviation of each Effect Size $g$ is: $V_g = J^2 \times V_d$ and $SE_g = \sqrt{V_g}$

\noindent where $V_d$ is the variance of Cohen's $d$ .Here ,again $n_1$,$n_2$ are the number of studies of control group and experimental group respectively:
\begin{equation}
\label{eq_Vd}
V_{d}=\frac{n_{1}+n_{2}}{n_{1} n_{2}}+\frac{d^{2}}{2\left(n_{1}+n_{2}\right)}
\end{equation}

\noindent Finally, each study's weight is calculated as: $W_{i}^{*} = \frac{1}{V_{g_i}^{*}} $ and $V_{g_i}^{*} = V_{g_i} + T^2$

where $V_{g_i}$ is the within-study variance for a study i and $T^2$ is the between-studies variance. In this study, $\tau^2$ is estimated with $T^2$ using the DerSimonian and Laird method \cite{dersimonian1986meta}

\section{Results and Discussions}
\label{sec:results}
The results section is discussing the detailed analysis of the discovered literature, then the answers to the research questions, and at the end of the section we discussed the meta-analytical results, presenting the meta-data and forest-plots.

\subsection{Literature Review}
We found 62 studies, the repository representation is as follows, 47.37\% from PROMISE, 42.10\% from NASA and 10.53\% from OSS. The answers to the research questions are as follows:

\begin{enumerate}[start=1, label={\bfseries RQ\arabic*:}, wide = 0pt, leftmargin = 3em]
\item Which repositories are popular for software defect prediction? \\
There were two types of repositories, public and private, in most of the studies the three repositories were discussed, PROMISE, NASA being public and OSS being private. When we analyzed the results, we found out PROMISE, NASA are commonly used, because of public availability.

\item What kinds of datasets are the most commonly utilized in prediction of software defects? \\
It can be visualized from the fig [fig.\ref{figure:datasetscommon}], that the researchers mostly used PC3, KCI, PC1, PC4. There is a marginal difference between the usage dataset, also the datasets are used in a combination with other datasets, but we can see that jEdit, KC3 and ANT were least commonly used.  

\item What are the programming languages which commonly in datasets for software defect prediction? \\
In the studies we observed that there were only four languages used, while Java and C were most common, whereas Perl was the least used Fig.\ref{figure:languags}. 

\begin{figure}[H]
\centering
\begin{minipage}[b][5cm][s]{.45\textwidth}
\centering
\caption{Most common datasets}
\vfill
\begin{tikzpicture}[scale = 0.70]
  \begin{axis}[
    ybar legend,
    legend style={at={(0.01,0.9)},anchor=west},
    ylabel={Count},
    symbolic x coords={PC3,KC1,PC1,PC4,CM1,JM1,PC2,ANT,KC3,jEdit},
    xtick=data,
    nodes near coords, 
    ymin=0,
    nodes near coords, style={xshift=0pt,yshift=0pt,anchor=north,font=\footnotesize},
    nodes near coords align={vertical},
    x tick label style={rotate=45,anchor=east},
    xlabel near ticks, 
    extra x tick style={xticklabel style={yshift=-15pt}},
    enlarge x limits , ybar=-0.35cm
    ]
    
    \addplot [black,fill=black!50!white]  coordinates {(PC3,10)(KC1,9)(PC1,9)(PC4,9)(CM1,8)(JM1,8)(PC2,7)(ANT,6)(KC3,6)(jEdit,5)};
     \end{axis}
\end{tikzpicture}
\vfill
\label{figure:datasetscommon}
\vspace{\baselineskip}
\end{minipage}\qquad
\begin{minipage}[b][5cm][s]{.45\textwidth}
\centering
\caption{Dataset languages}
\vfill
\begin{tikzpicture}[scale = .70]
  \begin{axis}[
    ybar legend,
    legend style={at={(0.01,0.9)},anchor=west},
    ylabel={Count},
    symbolic x coords={Java,C,C++,Perl},
    xtick=data,
    nodes near coords, 
    ymin=0,
    nodes near coords, style={xshift=0pt,yshift=0pt,anchor=north,font=\footnotesize},
    nodes near coords align={vertical},
    x tick label style={rotate=45,anchor=east},
    xlabel near ticks, 
    extra x tick style={xticklabel style={yshift=-15pt}},
    enlarge x limits , ybar=-0.35cm,
    bar width =7mm
    ]
    
    \addplot [black,fill=black!50!white]  coordinates {(Java,42)(C,39)(C++,17)(Perl,1)};
    
  \end{axis}
\end{tikzpicture}
\vfill
\label{figure:languags}
\vspace{\baselineskip}
\end{minipage}\qquad
\end{figure}

\item What are the common methods that are used to predict software defects? \\
As our research query only pinpoints the Naive Bayes and Random Forest, so these were the most common, besides these models we found out that Linear Regression and Decision Tree are also commonly used and Multi-layer perceptron was least used in Fig.\ref{figure:methodss}.

\item What comparison metrics are commonly used for comparing the software defect prediction performance? \\
The top discussed metrics are Accuracy, F-measure, Recall, and precision, one of the reason of these being top was the search query, and it is a known fact that accuracy is the common metric when we discuss ML models. Whereas, it was interesting to observe that precision was used fewer times, the least used metrics were MAE and ROC (Figure.\ref{figure:metricss}). 

\begin{figure}[H]
\centering
\begin{minipage}[b][5.1cm][s]{.45\textwidth}
\caption{Common Metrics}
\vfill
\begin{tikzpicture}[scale = .7]
  \begin{axis}[
    ybar legend,
    legend style={at={(0.01,0.9)},anchor=west},
    ylabel={Count},
    symbolic x coords={Accuracy,F-measure,Recall,Precision,AUC,MCC,MAE,ROC},
    xtick=data,
    nodes near coords, 
    ymin=0,
    nodes near coords, style={xshift=0pt,yshift=0pt,anchor=north,font=\footnotesize},
    nodes near coords align={vertical},
    x tick label style={rotate=45,anchor=east},
    xlabel near ticks, 
    extra x tick style={xticklabel style={yshift=-15pt}},
    enlarge x limits , ybar=-0.35cm
    ]
    \addplot [black,fill=black!50!white]  coordinates {(Accuracy,9)(F-measure,9)(Recall,8)(Precision,7)(AUC,6)(MCC,4)(MAE,2)(ROC,2)};
    
  \end{axis}
\end{tikzpicture}
\vfill
\label{figure:metricss}
\vspace{\baselineskip}
\end{minipage}\qquad
\begin{minipage}[b][5.1cm][s]{.45\textwidth}
\centering
\caption{Common Methods}
\vfill
\begin{tikzpicture}[scale = .7]
  \begin{axis}[
    ybar legend,
    legend style={at={(0.01,0.9)},anchor=west},
    ylabel={Count},
    symbolic x coords={NB,RF,LR,DT,SVM,K-NN,J48,MLP},
    xtick=data,
    nodes near coords, 
    ymin=0,
    nodes near coords, style={xshift=0pt,yshift=0pt,anchor=north,font=\footnotesize},
    nodes near coords align={vertical},
    x tick label style={rotate=45,anchor=east},
    xlabel near ticks, 
    extra x tick style={xticklabel style={yshift=-15pt}},
    enlarge x limits , ybar=-0.35cm
    ]
    
    \addplot [black,fill=black!50!white]  coordinates {(NB,15)(RF,12)(LR,7)(DT,7)(SVM,6)(K-NN,6)(J48,5)(MLP,4)};
    
  \end{axis}
\end{tikzpicture}
\vfill
\label{figure:methodss}
\vspace{\baselineskip}
\end{minipage}\qquad
\end{figure}
\end{enumerate}

\subsection{Meta-Analysis}
\label{meta_analysis_plots}
Meta-analysis is performed on the data we retrieved after literature review, the studies selected for meta-analysis are presented in Table \ref{tab:Meta-Analysis}. In the table, NB experiments represents the number of experiments performed for Naive Bayes, and RF for Random Forest. Y represents the presence of a specific measure in the study, and N means absence.
\begin{table*}[htb]
\centering
\begin{tabular}{|c|c|c|c|c|c|c|c|l|}
\hline
Title & Year & \begin{tabular}[c]{@{}c@{}}NB \\ Experiments\end{tabular} & \begin{tabular}[c]{@{}c@{}}RF \\ Experiments\end{tabular} & Recall & Precision & \begin{tabular}[c]{@{}c@{}}F-\\ Measure\end{tabular} & \begin{tabular}[c]{@{}c@{}}Total\\ Models\end{tabular} & \begin{tabular}[c]{@{}l@{}}Total\\ Metrics\end{tabular} \\ \hline
S1 \cite{alazzam2017software}    & 2017 & 4                                                         & 4                                                         & Y      & Y         & Y                                                    & 11                                                          & 4                                                                    \\ \hline
S2 \cite{kakkar2018os}    & 2018 & 7                                                         & 7                                                         & Y      & Y         & Y                                                    & 6                                                           & 5                                                                    \\ \hline
S3 \cite{david2018sdp}    & 2018 & 3                                                         & 3                                                         & N      & N         & Y                                                    & 4                                                           & 2                                                                    \\ \hline
S4 \cite{iqbal20198performance}    & 2019 & 24                                                        & 24                                                        & Y      & Y         & Y                                                    & 10                                                          & 6                                                                    \\ \hline
S5 \cite{khakhar2020integrity}    & 2020 & 3                                                         & 3                                                         & Y      & N         & N                                                    & 5                                                           & 2                                                                    \\ \hline
\end{tabular}
\caption{Meta-Data}
\label{tab:Meta-Analysis}
\end{table*}

Further, we drew forest plots to analyze the meta-data. Forest plot consists of 10 columns, where each row represents a study, the authors are the identifiers. Total is the number of experiments in the study, and mean and SD refer to the mean and standard deviation of that performance measure. The Standardized Mean Difference refers to the standardized difference between the Mean performance of both groups, and the values are in SMD column. The positions of the shaded rectangles represent the standardized mean differences for a study instance, while their size represents the effect weights. 

\begin{figure}[!ht]
      \centering
      \includegraphics[width=0.9\columnwidth]{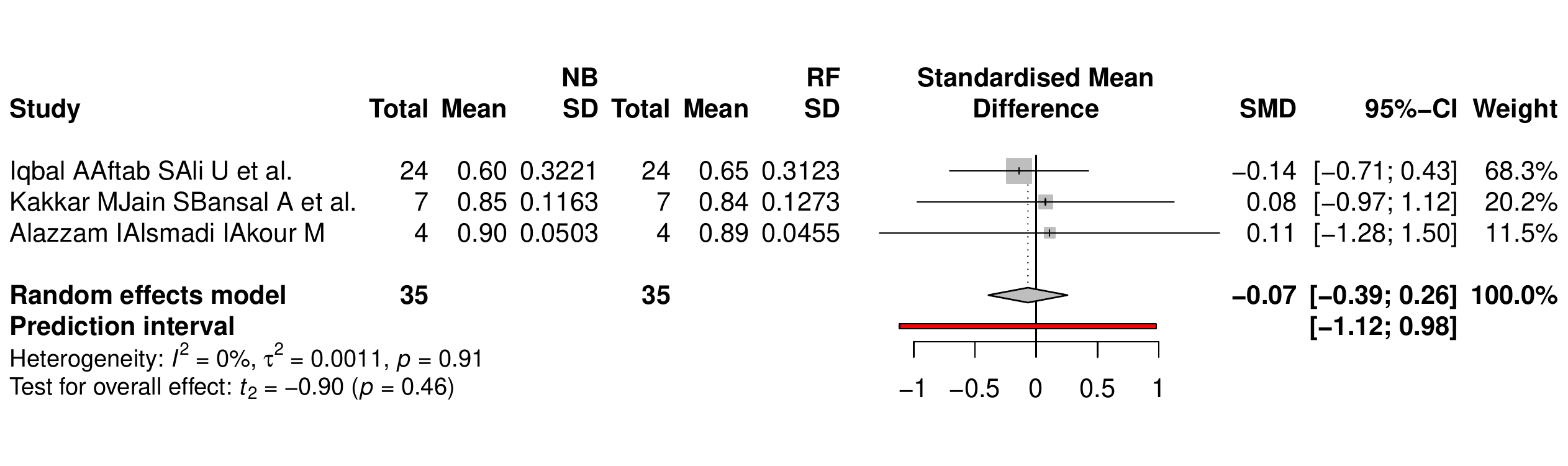}
      \caption{Forest plot for precision}
      \label{fig:plot_precision}
\end{figure}

\begin{figure}[!ht]
      \centering
      \includegraphics[width=0.9\columnwidth]{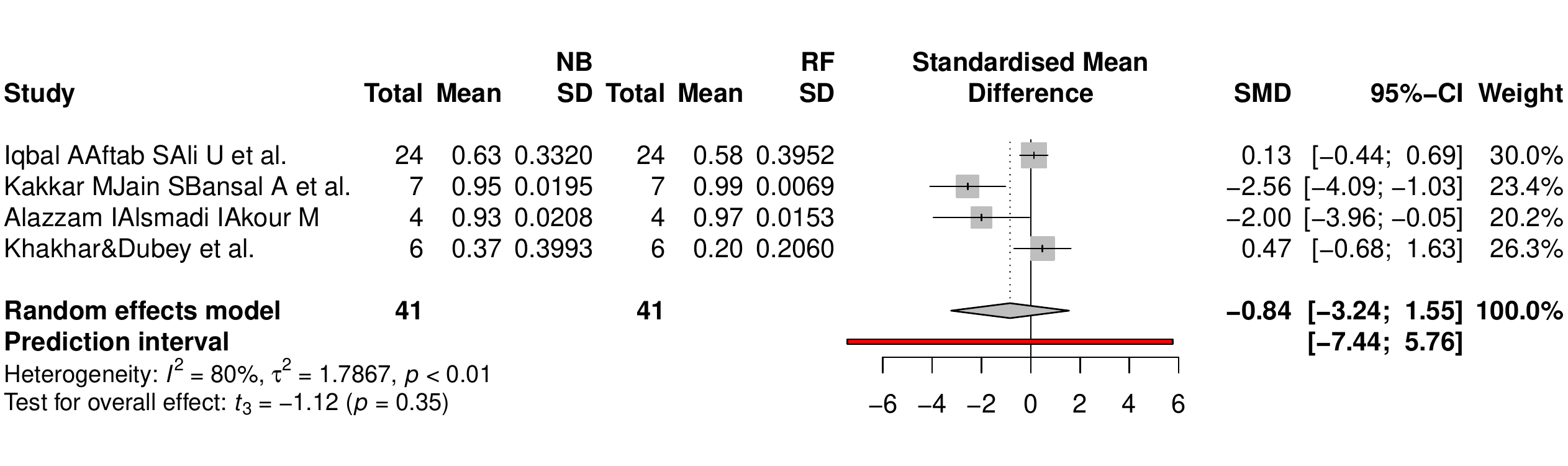}
      \caption{Forest plot for recall}
      \label{fig:plot_recall}
\end{figure}

\begin{figure}[!ht]
      \centering
      \includegraphics[width=0.9\columnwidth]{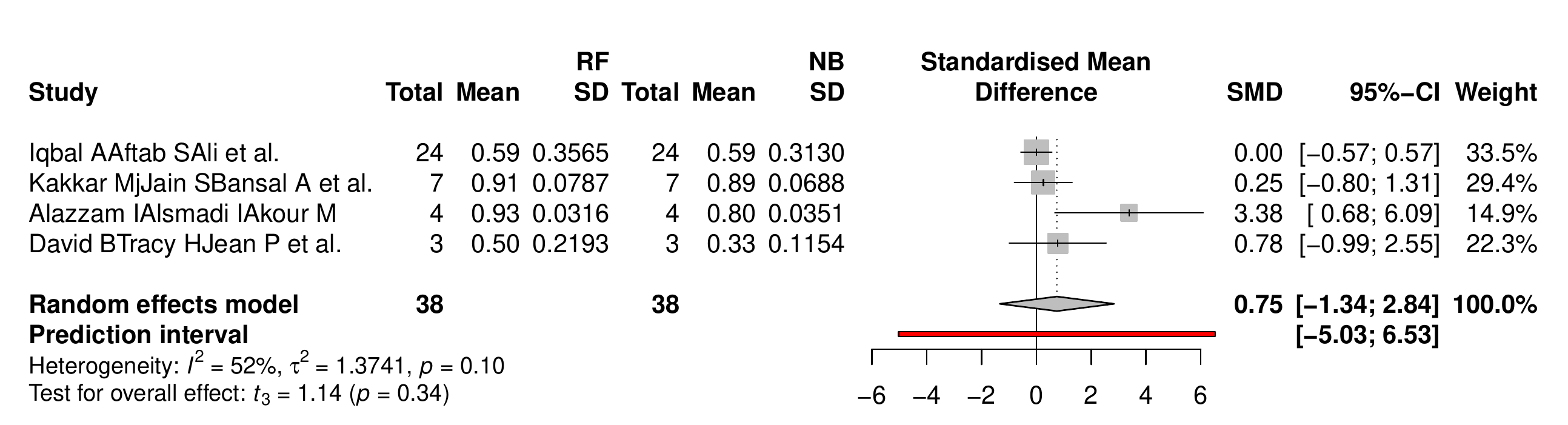}
      \caption{Forest plot for F-measure}
      \label{fig:plot_F_measure}
\end{figure}

95\% CI represents the confidence interval, verifies the hypothesis for SMD. Weights are the contribution of the study to the meta-analysis. Diamond represents the combined effect of all the studies. Heterogeneity identifies the model (random or fixed effect). Test for overall effect shows overall effect based on degrees of freedom. The forest plots are represented in Fig.\ref{fig:plot_precision}, Fig.\ref{fig:plot_recall} and Fig.\ref{fig:plot_F_measure}.


The standardized mean difference in the study is represented by the horizontal line of a forest plot. When the horizontal and vertical lines of a forest plot intersect, it means that there is no statistically significant difference between the research groups. These horizontal lines create a diamond, that symbolizes the combined impact of all the studies. There is no statistically significant difference between the groups for the studies when the diamond intersects the vertical line. From the forest-plots above, we can observe that all the diamonds are intersecting the vertical line, meaning that, the effects of Naive Bayes and Random Forests on software defect prediction are the same. The implementation details are publicly available in Github\footnote{\url{https://github.com/cm-awais/SDP_NB_RF_meta_analysis}}.

\section{Conclusions}
\label{sec:conclusion}
In this study, we thoroughly reviewed  literature assessment and performed a meta-analysis to determine if naive Bayes and random forest are more effective at predicting software problems. We collected the related studies, and presented statistics about databases, datasets, languages, repositories, machine learning models, comparison metrics. We retrieved 62 studies that address Naive Bayes and Random Forest, and five of them were chosen for meta-analysis based on our criteria. We conclude from the meta-analysis that there is no statistically significant difference between random forest and naive Bayes in terms of recall, precision, and f-measure for software defect prediction.

\section*{Acknowledgements}
This research is supported by the Russian Science Foundation, Grant No. 22-21-00494

\printbibliography

@article{zaugg2011mendeley,
  title={Mendeley: Creating communities of scholarly inquiry through research collaboration},
  author={Zaugg, Holt and West, Richard E and Tateishi, Isaku and Randall, Daniel L},
  journal={TechTrends},
  % volume={55},
  % number={1},
  % pages={32--36},
  year={2011},
  % publisher={Springer}
}

@article{j2009develop,
  title={Developing qualitative research questions: a reflective process},
  author={J Agee},
  journal={International Journal of Qualitative Studies in Education},
  % volume={22},
  % number={1},
  % pages={431--447},
  year={2009},
  % publisher={Taylor & Francis}
}

@article{he2015simplmetricSet,
  title={An empirical study on software defect prediction with a simplified metric set},
  author={He Peng, Li Bing},
  journal={Information and Software Technology},
  % volume={59},
  % number={1},
  % pages={170-190},
  year={2015},
  % publisher={Springer}
}

@article{son2019depMappings,
  title={Empirical study of software defect prediction: A systematic mapping},
  author={Son Le Hoang, Pritam Nakul, Khari Manju},
  journal={Symmetry},
  % volume={11},
  % number={1},
  % pages={32--36},
  year={2019},
  % publisher={Springer}
}

@article{martin2016researcherBias,
  title={Researcher Bias: The Use of Machine Learning in Software Defect Prediction},
  author={Martin Shepperd, David Bowes and Tracy Hall},
  % journal={IEEE Transactions on Software Engineering},
  % volume={42},
  % number={1},
  % pages={1092-1094},
  year={2016},
  % publisher={Springer}
}

@article{li2012sample,
  title={Sample-based software defect prediction with active and semi-supervised learning},
  author={Li, Ming and Zhang, Hongyu and Wu, Rongxin and Zhou, Zhi-Hua},
  journal={Automated Software Engineering},
  % volume={19},
  % number={2},
  % pages={201--230},
  year={2012},
  % publisher={Springer}
}

@article{kumar2016impacts,
  title={The impacts of test automation on software's cost, quality and time to market},
  author={Kumar, Divya and Mishra, KK},
  % journal={Procedia Computer Science},
  % volume={79},
  % pages={8--15},
  year={2016},
  % publisher={Elsevier}
}

@article{miller2000applying,
  title={Applying meta-analytical procedures to software engineering experiments},
  author={Miller, James},
  journal={Journal of Systems and Software},
  % volume={54},
  % number={1},
  % pages={29--39},
  year={2000},
  % publisher={Elsevier}
}

@article{aleem2015benchmarking,
  title={Benchmarking machine learning technologies for software defect detection},
  author={Aleem, Saiqa and Capretz, Luiz Fernando and Ahmed, Faheem},
  % journal={arXiv preprint arXiv:1506.07563},
  year={2015}
}

@inproceedings{devnani1998modeling,
  title={Modeling software defect introduction},
  author={Devnani-Chulani, Sunita},
  booktitle={Proc. California Software Symposium},
  % volume={97},
  year={1998}
}

@book{lewis2017software,
  title={Software testing and continuous quality improvement},
  author={Lewis, William E},
  year={2017},
  publisher={CRC press}
}

@book{zuse2019software,
  title={Software complexity: measures and methods},
  author={Zuse, Horst},
  % volume={4},
  year={2019},
  publisher={Walter de Gruyter GmbH \& Co KG}
}

@article{li2018progress,
  title={Progress on approaches to software defect prediction},
  author={Li, Zhiqiang and Jing, Xiao-Yuan and Zhu, Xiaoke},
  journal={IET Software},
  % volume={12},
  % number={3},
  % pages={161--175},
  year={2018},
  % publisher={IET}
}

@article{jacob2015improved,
  title={Improved random forest algorithm for software defect prediction through data mining techniques},
  author={Jacob, Shomona Gracia and others},
  journal={IJCA},
  % volume={117},
  % number={23},
  year={2015},
  % publisher={Citeseer}
}

@article{petticrew2008picoc,
  title={Systematic reviews in the social sciences: A practical guide},
  author={M Petticrew, H Roberts},
  journal={Acp j club},
  % volume={123},
  % number={3},
  % pages={43--44},
  year={2008}
}

@article{leandro2008meta,
  title={Meta-analysis in medical research: The handbook for the understanding and practice of meta-analysis},
  author={G Leandro},
  journal={Acp j club},
  % volume={123},
  % number={3},
  % pages={43--44},
  % link = {https://books.google.com/books?hl=en&lr=&id=kNeSIHNbJs4C&oi=fnd&pg=PR7&dq=+meta+analysis+software&ots=jQ-OuCrHNu&sig=6BgIS8ayM5HcchDqI9e8lU9OFow},
  year={2008}
}

@inproceedings{soe2018NB,
  title={A Comparison of Na{\"\i}ve Bayes and Random Forest for Software Defect Prediction},
  author={Soe, Yan Naung and Oo, Khine Khine},
  year={2018},
  organization={ICCA}
}

@article{venteka2005emperical,
  title={Empirical Assessment of Machine Learning based Software Defect Prediction Techniques},
  author={Venkata U.B. Challagulla, },
  journal={Acp j club},
  % volume={123},
  % number={3},
  % pages={43--44},
  % link = {https://books.google.com/books?hl=en&lr=&id=kNeSIHNbJs4C&oi=fnd&pg=PR7&dq=+meta+analysis+software&ots=jQ-OuCrHNu&sig=6BgIS8ayM5HcchDqI9e8lU9OFow},
  year={2005}
}

@inbook{intro_to_Meta_book_2009.ch16,
    % publisher = {John Wiley \& Sons, Ltd},
    % isbn = {9780470743386},
    title = {Identifying and Quantifying Heterogeneity},
    booktitle={Introduction to meta-analysis},
    author={Borenstein, Michael and Hedges, Larry V and Higgins, Julian PT and Rothstein, Hannah R},
    chapter = {16},
    pages = {107-125},
    % doi = {https://doi.org/10.1002/9780470743386.ch16},
    % url = {https://onlinelibrary.wiley.com/doi/abs/10.1002/9780470743386.ch16},
    % eprint = {https://onlinelibrary.wiley.com/doi/pdf/10.1002/9780470743386.ch16},
    year = {2009}
}

@inbook{c2009systematic.ch1,
    % publisher = {CRD, University of York},
    % isbn = {978-1-900640-47-3},
    title = {CRD’s guidance for undertaking reviews in health care},
    booktitle={Introduction to meta-analysis},
    author={University of York},
    chapter = {1},
    pages = {54-55},
    % doi = {https://doi.org/10.1002/9780470743386.ch16},
    % url = {https://www.york.ac.uk/media/crd/Systematic_Reviews.pdf},
    year = {2009}
}

@article{iqbal20198performance,
  title={Performance analysis of machine learning techniques on software defect prediction using NASA datasets},
  author={A Iqbal, S Aftab, U Ali},
  % journal={IJACSA},
  % volume={10},
  % number={5},
  % pages={300--308},
  year={2019}
}

@article{khakhar2020integrity,
  title={The integrity of machine learning algorithms against software defect prediction},
  author={P. Khakhar, R. Dubey},
  % journal={arXiv},
  % volume={123},
  % number={3},
  % pages={1--7},
  year={2020}
}

@article{alazzam2017software,
  title={Software fault proneness prediction: a comparative study between bagging, boosting, and stacking ensemble and base learner methods},
  author={I. Alazzam, I. Alsmadi, M. Akour},
  % journal={IJDATS},
  % volume={9},
  % number={1},
  % pages={1--1},
  year={2017}
}

@article{kakkar2018os,
  title={Is Open-Source Software Valuable for Software Defect Prediction of Proprietary Software and Vice Versa?},
  author={M. Kakkar, S. Jain},
  % journal={Advances in Intelligent Systems and Computing},
  % volume={589},
  % number={3},
  % pages={227--236},
  year={2018}
}

@article{dersimonian1986meta,
  title={Meta-analysis in clinical trials},
  author={DerSimonian, Rebecca and Laird, Nan},
  journal={Controlled clinical trials},
  % volume={7},
  % number={3},
  % pages={177--188},
  year={1986},
  % publisher={Elsevier}
}

@article{david2018sdp,
  title={Software defect prediction: do different classifiers find the same defects?},
  author={David Bowes, Tracy Hall, Jean Petric},
  journal={Software Qual J},
  % volume={26},
  % number={3},
  % pages={525-552},
  % link = {https://link.springer.com/article/10.1007/s11219-016-9353-3},
  year={2018}
}

\end{document}